\documentclass[amsmath,amssymb,amsfonts,aps,pre,preprint,superscriptaddress,bibnotes,showpacs,showkeys,longbibliography]{revtex4-1}

\usepackage{graphicx}
\usepackage{subfigure}
\usepackage{amsmath}
\usepackage{amsfonts}
\usepackage{amssymb}
\usepackage{natbib}
\usepackage{relsize}
\usepackage{sidecap}
\usepackage{braket}
\usepackage{footmisc}
\usepackage{footnote}
\usepackage{color}
\begin{document}

\title{Casimir effect in a dilute Bose gas in canonical ensemble within improved Hartree-Fock approximation}

\author{Nguyen Van Thu}
\affiliation{Department of Physics, Hanoi Pedagogical University 2, Hanoi 100000, Vietnam}
\affiliation{Institute of Research and Development, Duy Tan University, Danang 550000, Vietnam}

\begin{abstract}
The Casimir effect in a dilute Bose gas confined between two planar walls is investigated in the canonical ensemble at zero temperature by means of Cornwall-Jackiw-Tomboulis effective action approach within the improved Hartree-Fock approximation. Our results show that: (i) the Casimir energy and the resulting Casimir force in the canonical ensemble remarkably differ from those in the grand canonical ensemble; (ii) when the distance between two planar walls increases, the Casimir energy and Casimir force decay in accordance with a half-integer power law in the canonical ensemble instead of an integer power law in the grand canonical ensemble.
\end{abstract}

\maketitle

\section{Introduction\label{sec1}}

More than seventy years have passed since the Casimir effect was discovered \cite{Casimir}, this effect is and will be a widely-mentioned subject of modern physics in many areas, for example, quark matter \cite{ThuPhat}, atomic and molecular physics \cite{Babb}, superconductor \cite{Bimonte,Chen} and so on in both the theory and experiment. In field of Bose-Einstein condensate (BEC), motivated by experiments on the critical Casimir force \cite{Fukuto,Ganshin} and the Casimir-Polder force \cite{Harber,Obrecht,Klimchitskaya}, the interest in this phenomenon has blossomed in recent years.

It is well-known that, in quantum theory, the Casimir effect in the BEC at zero temperature confined by two planar walls is the consequence of the quantum fluctuation above ground state associated with phononic excitations \cite{Biswas2,Biswas3,Schiefele}. So far, many studies on the Casimir effect in the BEC mainly focused on considering it at zero temperature within the constraint of a dilute condition. In process of the calculations, one has to deal with the ultraviolet (UV) divergence of the energy density.  To eliminate this problem, some methods are available, such as, using Abel-Plana formula as in \cite{Schiefele,Thu1}, momentum cut-off and using Euler-MacLaurin formula \cite{Biswas3,Pomeau}. For the weakly interacting Bose gas, in Ref. \cite{Schiefele,Thu1} the Casimir force is expressed through an integral of density of state. Instead of using Abel-Plana formula, using the second way, in the one-loop approximation of quantum field theory, many authors pointed out that the Casimir force in a weakly interacting Bose gas depends on the coupling constant and decays as power law of the distance $\ell$ between two planar walls, which is the well-known result
\begin{eqnarray}
F_C=-\frac{\pi^2}{480}\frac{\hbar v_s}{\ell^4},\label{1a}
\end{eqnarray}
in which $\hbar$ is Planck's constant, the speed of sound $v_s$ is determined by the interaction of atoms in the system.

In the high-order approximation, our previous work \cite{ThuIJMPB} employed the Cornwall-Jackiw-Tomboulis (CJT) effective action approach \cite{CJT} to consider the Casimir effect in the improved Hartree-Fock (IHF) approximation in the lowest order of the momentum integrals. Our results show that although the order parameter is strongly affected by the compactification of the space, the Casimir energy and thus the Casimir force are the same in comparison  with those in the one-loop approximation. More exactly, this effect is considered in Ref. \cite{ThuPhysa} in the higher-order term of the momentum integrals and the obtained results are significantly improved.  The Casimir force equals to its value in the one-loop approximation (\ref{1a}) after adding a corrected term and
\begin{eqnarray}
F_C\propto-\frac{1}{\ell^4}-\frac{1}{\ell^7}.\label{1b}
\end{eqnarray}
Eqs. (\ref{1a}) and (\ref{1b}) show that the Casimir force decays according to the law of the negative half-integer power of the distance.

A common feature of the above studies is that the system under consideration is connected to a particle reservoir, this means that the results were given in the grand canonical ensemble. From the idea of Fisher and de Gennes \cite{Fisher} about the critical Casimir effect, the authors of Refs. \cite{Gross,Rohwer} investigated it in the critical fluid system within the mean-field theory in the canonical ensemble, where the particle number is kept constant. In these works, it has been found that significant qualitative differences arise between the critical Casimir force in the canonical ensemble and the grand canonical ensemble. These studies authors also gave evidences that the critical Casimir force strongly depends on the imposed boundary conditions. To our knowledge, there is a lack of investigating the Casimir effect in the BEC at zero temperature in the canonical ensemble, except our work \cite{Thu382}. However, the Casimir effect was only considered in the one-loop approximation. To fill this gap and improve the accuracy of the results, in this paper, the Casimir effect is worked out in the canonical ensemble within the framework of the CJT effective action approach in the IHF approximation, in which not only the contribution of all two-loop diagrams is taken into account, but the Goldstone theorem is also valid.

To begin with, we consider a dilute Bose gas described by Lagrangian \cite{Pethick},
\begin{eqnarray}
{\cal L}=\psi^*\left(-i\hbar\frac{\partial}{\partial t}-\frac{\hbar^2}{2m}\nabla^2\right)\psi-\mu\left|\psi\right|^2+\frac{g}{2}\left|\psi\right|^4,\label{1}
\end{eqnarray}
with $\psi=\psi(\vec{r},t)$ being the field operator, its expectation value plays the role of the order parameter; $m$ and $\mu$ are the atomic mass and chemical potential, respectively; the strength of repulsive intraspecies interaction is determined by coupling constant $g=4\pi\hbar^2a_s/m>0$ with $a_s$ being the $s$-wave scattering length \cite{Pitaevskii}.

This paper is structured as follows. In Section \ref{sec:2}, the Casimir effect in a dilute Bose gas is studied in the canonical ensemble, which consists of the Casimir energy and the Casimir force. Conclusion and outlook are given in Section \ref{sec:3}.

\section{Casimir effect in a dilute Bose gas in canonical ensemble \label{sec:2}}

In this Section, the Casimir effect in a dilute Bose gas confined between two planar walls is considered in the canonical ensemble and in the improved Hartree-Fock approximation at zero temperature. At first, we consider a Bose gas at finite temperature $T$ and let the temperature be zero at the end. Let $\psi_0$ be the expectation value of the field operator $\psi$ in the mean field theory, an expansion of the field operator can be performed
\begin{eqnarray}
\psi\rightarrow \psi_0+\frac{1}{\sqrt{2}}(\psi_1+i\psi_2),\label{shift}
\end{eqnarray}
in which two real fields $\psi_1,\psi_2$ associated with the quantum fluctuations of the field \cite{Andersen}.
By substituting (\ref{shift}) into Lagrangian (\ref{1}), one has the free Lagrangian
\begin{eqnarray}
{\cal L}_0=-\mu\psi_0^2+\frac{g}{2}\psi_0^4,\label{L0}
\end{eqnarray}
and the interaction Lagrangian in the double-bubble approximation
\begin{eqnarray}
{\cal L}_{int}=\frac{g}{2}\psi_0\psi_1(\psi_1^2+\psi_2^2)+\frac{g}{8}(\psi_1^2+\psi_2^2)^2.\label{Lint}
\end{eqnarray}
Based on (\ref{L0}), it is easy to find the inverse propagator in the tree-approximation
\begin{eqnarray}
D_0^{-1}(k)&=&\left(
              \begin{array}{lr}
                \frac{\hbar^2\vec{k}^2}{2m}+2g\psi_0^2 & -\omega_n \\
                \omega_n &  \frac{\hbar^2\vec{k}^2}{2m}\\
              \end{array}
            \right),\label{protree}
\end{eqnarray}
in momentum space, where $\vec{k}$ is the wave vector. The Matsubara frequency for boson is defined as $\omega_n=2\pi n/\beta ~(n=0,\pm1,\pm2...),~\beta=1/k_BT$ and $k_B$ is Boltzmann constant. The inversion propagator (\ref{protree}) shows that there is a Goldstone boson associated with the symmetric breaking of $U(1)$ group.

It was pointed out \cite{ThuIJMPB,ThuPhysa,Phat} that in the Hartree-Fock approximation, the Goldstone boson is disappeared. To restore it, a method proposed by Ivanov {\it et. al.} in Ref. \cite{Ivanov} is invoked and which leads to an approximation called the IHF approximation. In this approximation, the CJT effective potential has the form
\begin{eqnarray}
\widetilde{V}_\beta^{CJT}=&&-\mu\psi_0^2+\frac{g}{2}\psi_0^4+\frac{1}{2}\int_\beta \mbox{tr}\left[\ln D^{-1}(k)+D_0^{-1}(k)D(k)-{1\!\!1}\right]\nonumber\\
&&+\frac{g}{8}(P_{11}^2+P_{22}^2)+\frac{3g}{8}P_{11}P_{22},\label{VIHF}
\end{eqnarray}
which corresponds to a new inverse propagator
\begin{eqnarray}
D^{-1}(k)=\left(
              \begin{array}{lr}
                \frac{\hbar^2k^2}{2m}+M^2 & -\omega_n \\
                \omega_n & \frac{\hbar^2k^2}{2m} \\
              \end{array}
            \right),\label{proIHF}
\end{eqnarray}
with $M$ being the effective mass of the Goldstone boson, which is restored in the IHF approximation. In Eqs. (\ref{VIHF}) and (\ref{proIHF}), we abbreviate
\begin{eqnarray*}
\int_\beta f(\vec{k})=\frac{1}{\beta}\sum_{n=-\infty}^{+\infty}\int\frac{d^3\vec{k}}{(2\pi)^3}f(\omega_n,\vec{k}).
\end{eqnarray*}
and
\begin{eqnarray}
&&P_{11}\equiv \int_\beta D_{11}(k)=\frac{1}{2}\int\frac{d^3\vec{k}}{(2\pi)^3}\sqrt{\frac{\hbar^2k^2/2m}{\hbar^2k^2/2m+M^2}},\nonumber\\
&&P_{22}\equiv \int_\beta D_{22}(k)=\frac{1}{2}\int\frac{d^3\vec{k}}{(2\pi)^3}\sqrt{\frac{\hbar^2k^2/2m+M^2}{\hbar^2k^2/2m}},\label{tichphan}
\end{eqnarray}
are momentum integrals at zero temperature. The dispersion relation in the IHF approximation can be read off by requesting the determinant of the inversion propagator vanishes \cite{Floerchinger}, which yields
\begin{eqnarray}
E(k)=\sqrt{\frac{\hbar^2k^2}{2m}\left(\frac{\hbar^2k^2}{2m}+M^2\right)}.\label{dispersion}
\end{eqnarray}

Using the formula \cite{Schmitt},
\begin{eqnarray*}
\frac{1}{\beta}\sum_{n=-\infty}^{n=+\infty}\ln\left[\omega_n^2+E^2(k)\right]&=&E(k)+\frac{2}{\beta}\ln\left[1-e^{-\beta E(k)}\right],
\end{eqnarray*}
one can extract the canonical energy density at zero temperature from Eqs. (\ref{VIHF}) and (\ref{dispersion}),
\begin{eqnarray}
\Omega\equiv\frac{1}{2}\int_\beta \mbox{tr}\ln D^{-1}(k)=\frac{1}{2}\int\frac{d^3\vec{k}}{(2\pi)^3}\sqrt{\frac{\hbar^2k^2}{2m}\left(\frac{\hbar^2k^2}{2m}+M^2\right)}.\label{en}
\end{eqnarray}
In order to simplify notations, dimensionless quantities are introduced: wave vector $\kappa=k\xi$, effective mass ${\cal M}=M/\sqrt{gn_0}$ with $\xi=\hbar/\sqrt{2mgn_0}$ and $n_0$ being the healing length and bulk density, respectively. In this way, the momentum integrals in Eqs. (\ref{tichphan}) become
\begin{eqnarray}
&&P_{11}=\frac{1}{2\xi^3}\int\frac{d^3\kappa}{(2\pi)^3}\frac{\kappa}{\sqrt{\kappa^2+{\cal M}^2}},~P_{22}=\frac{1}{2\xi^3}\int\frac{d^3\kappa}{(2\pi)^3}\frac{\sqrt{\kappa^2+{\cal M}^2}}{\kappa},\label{tichphan1a}
\end{eqnarray}
and the canonical energy density (\ref{en}) can be rewritten
\begin{eqnarray}
\Omega=\frac{gn_{0}}{2\xi^3}\int\frac{d^3\kappa}{(2\pi)^3}\sqrt{\kappa^2(\kappa^2+{\cal M}^2)}.\label{energy}
\end{eqnarray}
The integrations over the dimensionless wave vector are UV-divergent. This divergence can be eliminated by using  the dimensional regularization proposed by Andersen in Ref. \cite{Andersen}.

Now the effects from the compactification of one direction in space on the Casimir energy and the Casimir force are considered. The system under consideration is the weakly interacting Bose gas confined by two planar walls, which are parallel plates at the distance $\ell$ and are perpendicular to $0z$-axis; the order parameter is translational invariant in the $0x$ and $0y$ directions. The area $A$ of each planar wall satisfies $A\gg\ell^2$. Our system is not connected to any particle reservoir therefore the particle number $N$ is fixed and roughly speaking
\begin{eqnarray}
N=n_0A\ell.\label{number}
\end{eqnarray}
This means that the system is considered in the canonical ensemble.

Because of this compactification, the wave vector is quantized as follows
\begin{eqnarray*}
k^2\rightarrow k_\perp^2+k_n^2,
\end{eqnarray*}
in which the wave vector component $k_\perp$ is perpendicular to $0z$-axis and $k_n$ is parallel with $0z$-axis. For the  boson system, the periodic boundary condition is imposed
\begin{eqnarray*}
k_n=\frac{2\pi n}{\ell}.
\end{eqnarray*}
In dimensionless form one has
\begin{eqnarray}
\kappa^2\rightarrow \kappa_\perp^2+\kappa_n^2, ~\kappa_{n}=\frac{2\pi n}{L}\equiv\frac{n}{\overline{L}},~\overline{L}=\frac{L}{2\pi},\label{k1}
\end{eqnarray}
where $L=\ell/\xi$. The quantization of wave vector leads to a change in the canonical energy density and Eq. (\ref{energy}) becomes
\begin{eqnarray}
\Omega=\frac{8 \pi ^3\hbar ^2Na_s}{mA \ell^3 }\sum_{n=-\infty}^\infty\int\frac{d^2\kappa_{\perp}}{(2\pi)^2}\sqrt{(\overline{L}^2\kappa_{\perp}^2+n^2)({\cal M}_0^2+n^2)},\label{term12}
\end{eqnarray}
in which
\begin{eqnarray*}
{\cal M}_0=\overline{L}\sqrt{\kappa_{\perp}^2+{\cal M}^2}.
\end{eqnarray*}
It is easy to check that the integral in Eq. (\ref{term12}) is UV-divergent when $\kappa_\perp$ tends to infinity, thus a cut-off $\Lambda$ is introduced for the top limit of this integral. After integrating over $\kappa_\perp$ from zero to $\Lambda$, the summation in Eq. (\ref{term12}) can be dealt by using the Euler-Maclaurin formula \cite{Arfken},
\begin{eqnarray}
\sum_{n=0}^\infty \theta_nF(n)-\int_0^\infty F(n)dn=-\frac{1}{12}F'(0)+\frac{1}{720}F'''(0)-\frac{1}{30240}F^{(5)}(0)+\cdots,\label{EM}
\end{eqnarray}
with
\begin{eqnarray*}
\theta_n=\left\{
           \begin{array}{ll}
             1/2, & \hbox{if $n=0$;} \\
             1, & \hbox{if $n>0$.}
           \end{array}
         \right.
\end{eqnarray*}
Finally, after taking the limit $\Lambda\rightarrow\infty$ and subtracting the bulk part one has a finite part of the canonical energy density
\begin{eqnarray}
{\cal E}_C=-\frac{\pi^{5/2}\hbar^2a_s^{1/2}}{720\sqrt{2}m}\sqrt{\frac{N}{A}}\frac{{\cal M}}{\ell^{7/2}},\label{energyC}
\end{eqnarray}
which is the Casimir energy. Eq. (\ref{energyC}) makes it clear that the Casimir energy in the canonical ensemble is proportional to $\ell^{-7/2}$ instead of $\ell^{-3}$ in the grand canonical ensemble \cite{ThuPhysa}.

The momentum integrals (\ref{tichphan1a}) formally resemble the canonical energy density (\ref{term12}). Under the quantization of the wave vector, the momentum integrals (\ref{tichphan1a}) are also UV-divergent and it can be resolved in the same way as the method applied for the Casimir energy. By introducing the cut-off $\Lambda$, Eqs. (\ref{tichphan1a}) become
\begin{eqnarray}
P_{11}&=&8\sqrt{2}\left(\frac{Na_s}{A\ell}\right)^{3/2}\sum_{n=-\infty}^{+\infty}\int_0^\Lambda \frac{d^2\kappa_\perp}{(2\pi)^2}\sqrt{\frac{\kappa_\perp^2+\kappa_n^2}{\kappa_\perp^2+\kappa_n^2+{\cal M}^2}},\nonumber\\
P_{22}&=&8\sqrt{2}\left(\frac{Na_s}{A\ell}\right)^{3/2}\sum_{n=-\infty}^{+\infty}\int_0^\Lambda \frac{d^2\kappa_\perp}{(2\pi)^2}\sqrt{\frac{\kappa_\perp^2+\kappa_n^2+{\cal M}^2}{\kappa_\perp^2+\kappa_n^2}}, \label{k2}
\end{eqnarray}
and using Euler-Maclaurin formula (\ref{EM}) one has
\begin{eqnarray}
P_{11}=-\frac{\pi ^2}{180 \ell^3 \cal M}, ~P_{22}=\frac{\pi N a_s \cal M}{3 A \ell^2}-\frac{\pi ^2}{180 \ell^3 \cal M}.\label{k4}
\end{eqnarray}

In order to calculate the Casimir force, the remaining problem is finding the dimensionless effective mass $\cal M$ in Eq. (\ref{energyC}). To this end, we first note that for a dilute Bose gas, i.e. the gas parameter must satisfy the well-known condition $n_0a_s^3\ll1$ \cite{Pethick}. In this case, the chemical potential is approximated $\mu\approx gn_0$ \cite{Andersen}. Minimizing the CJT effective potential (\ref{VIHF}) with respect to the order parameter and elements of the propagator, keeping in mind Eqs. (\ref{k4}), one arrives at the dimensionless form of the gap equation
\begin{eqnarray}
-1+\phi_0^2+\frac{\pi a_s\cal M}{6 \ell}-\frac{\pi ^2 A}{90N{\cal M} \ell^2}=0,\label{gap}
\end{eqnarray}
and the Schwinger-Dyson equation
\begin{eqnarray}
-1+3\phi_0^2+\frac{\pi a_s\cal M}{2\ell}-\frac{\pi ^2 A}{90N{\cal M} \ell^2}={\cal M}^2.\label{SD}
\end{eqnarray}
Note that here we use $\phi_0=\psi_0/\sqrt{n_0}$. It is easy to prove that the solution for Eqs. (\ref{gap}) and (\ref{SD}) has the form
\begin{eqnarray}
{\cal M}=\sqrt{\frac{2}{3}}\cos\frac{\alpha}{3},\label{Mhdn}
\end{eqnarray}
in which
\begin{eqnarray}
\cos\alpha=\frac{\pi ^2 A}{20 \sqrt{6}N \ell^2}.\label{alpha}
\end{eqnarray}
For the weakly interacting Bose gas, it is shown that \cite{ThuPhysa} the dimensionless effective mass can be written in a approximate form
\begin{eqnarray}
{\cal M}\approx{\cal M}_1+\frac{1}{3}\sqrt{\frac{2}{3}}\cos\alpha,\label{Mhd1}
\end{eqnarray}
in which ${\cal M}_1=\sqrt{2}$ is the dimensionless effective mass in the one-loop approximation. This result is an improvement compared with the one in the one-loop approximation and the second term in right hand side of Eq. (\ref{Mhd1}) is a result of taking into account the two-loop diagrams.

Combining Eqs. (\ref{energyC}) and (\ref{Mhd1}), the Casimir energy can be written as a function of only the distance between two planar walls
\begin{eqnarray}
{\cal E}_C=-\frac{\pi^{5/2}\hbar^2a_s^{1/2}}{720m}\sqrt{\frac{N}{A}}\frac{1}{\ell^{7/2}}-\frac{\pi^{9/2}\hbar^2a_s^{1/2}}{180.720\sqrt{2}m}\sqrt{\frac{A}{N}}\frac{1}{\ell^{11/2}}.\label{E1}
\end{eqnarray}
It is not difficult to realize that the first term in right hand side of Eq. (\ref{E1}) is the Casimir energy in the one-loop approximation in the canonical ensemble \cite{Thu382} because in this approximate level the dimensionless effective mass is independent of the distance as shown in Eq. (\ref{Mhd1}). The second term is contribution of the high-order diagram in the interaction Lagrangian (\ref{Lint}). The common feature can be read off from (\ref{E1}) is that the Casimir energy is negative in the canonical ensemble, this is similar to the one in the grand canonical ensemble \cite{ThuIJMPB,ThuPhysa}.

\begin{figure}[h]
\leavevmode
  \includegraphics[scale=0.8]{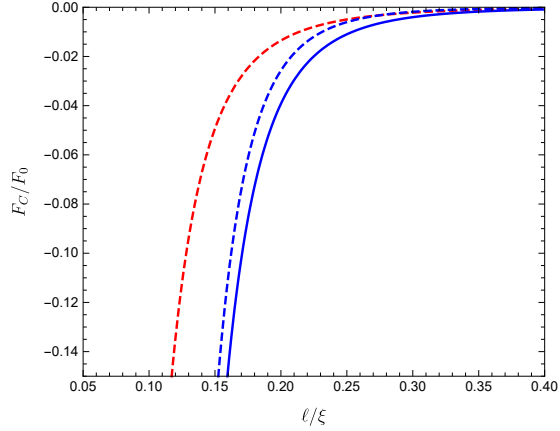}
  \caption{\footnotesize (Color online) The evolution of the Casimir force versus the distance (solid line). The red dashed line and the blue dashed line correspond to the first and second terms in right hand side of Eq. (\ref{Fc1}).}\label{f1}
\end{figure}

The Casimir force is defined as the negative derivative of the Casimir energy with respect to a change in the distance between two planar walls
\begin{eqnarray}
F_C=-\frac{\partial {\cal E}_C}{\partial\ell}.\label{dn}
\end{eqnarray}
Substituting (\ref{E1}) into (\ref{dn}) one has
\begin{eqnarray}
F_C=-\frac{7\pi^{5/2}\hbar^2a_s^{1/2}}{1440m}\sqrt{\frac{N}{A}}\frac{1}{\ell^{9/2}}-\frac{11\pi^{9/2}\hbar^2a_s^{1/2}}{180.1440m}\sqrt{\frac{A}{N}}\frac{1}{\ell^{13/2}}.\label{Fc}
\end{eqnarray}
As a consequence of the Casimir energy, the first term in right hand side of Eq. (\ref{Fc}) is the Casimir force in the  one-loop approximation in the canonical ensemble and the last one comes from the correction of the high-order diagrams in the interaction Lagrangian. Moreover, when looking at the Eq. (\ref{Fc}) there is no doubt that the Casimir force is attractive, the same as the one in the grand canonical ensemble. In dimensionless form, Eq. (\ref{Fc}) is scaled by
\begin{eqnarray}
F_0=\frac{64\sqrt{2}\pi^{5/2}\hbar^2}{m}\left(\frac{N}{A}\right)^{5/2},
\end{eqnarray}
and thus
\begin{eqnarray}
\frac{F_C}{F_0}=-\frac{7A^2a_s^{1/2}}{92160\sqrt{2}N^2}\frac{1}{\ell^{9/2}}-\frac{11\pi^2A^3a_s^{1/2}}{360.92160N^3}\frac{1}{\ell^{13/2}}.\label{Fc1}
\end{eqnarray}
In order to illustrate for above calculations, several numerical computations are made for rubidium Rb 87 with $m=86.909$ u (1 u = 1.6605.$10^{-27}$ kg is unit of atomic mass), $a_s=100.4a_0,~a_0=0.529$ {\AA} is Bohr radius \cite{Egorov}. The total particle number and the area of each planar wall are $N=6.10^6,~A=10^{-6}$ m$^2$ \cite{Biswas3}, respectively. The $\ell$-dependence of the Casimir force is shown in Fig. \ref{f1}, in which the dashed red and dashed blue lines correspond to the first and second term in right hand side of (\ref{Fc1}), the solid blue line sketches the Casimir force, in which the distance between two planar walls is scaled by $\xi=4000$ {\AA}. It is obvious that the Casimir is only significant in small-$\ell$ region and contribution of the second term is dominant.
\begin{figure}[h]
\leavevmode
  \includegraphics[scale=0.8]{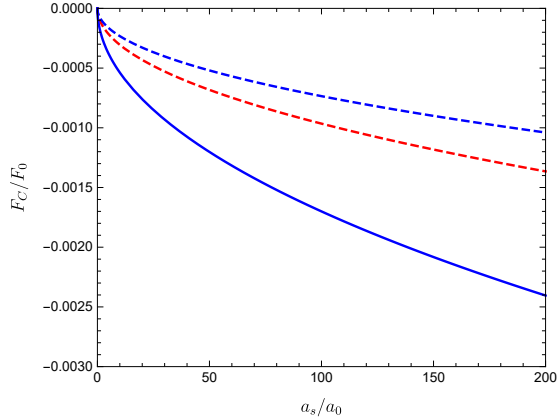}
  \caption{\footnotesize (Color online) The evolution of the Casimir force versus the $s$-wave scattering length (solid line). The red dashed line and the blue dashed line correspond to the first and second terms in right hand side of Eq. (\ref{Fc1}).}\label{f2}
\end{figure}

An important property of Bose gas is that the $s$-wave scattering length $a_s$ can be controlled by Feshbach resonance \cite{Inouye} changing the laser detuning. Fig. \ref{f2} clearly shows that the Casimir force against the scattering length $a_s$ at distance $\ell=4000$ {\AA} with the same parameters in Fig. \ref{f1}. In this case, the first and second terms contribute to the Casimir force in the same order. These lines are in shape of $a_s^{1/2}$ and thus the Casimir force tends to zero when the scattering length approaches to zero, this result coincides to the fact that the Casimir force in an ideal Bose gas vanishes at zero temperature because of vanishing of the speed of sound as pointed out in Eq. (\ref{1a}).

\section{Conclusion and Outlook\label{sec:3}}

We presented here the analysis of the Casimir effect in the dilute Bose gas confined between two parallel plates in the canonical ensemble in the improved Hartree-Fock approximation with the conservation of Goldstone boson. Our results indicate that Casimir force is significantly different not only in the different ensembles and the same approximation (the IHF approximation) but also in different approximations and the same ensemble (the canonical ensemble).

First of all, within the same approximation, i.e. the improved Hartree-Fock approximation, the Casimir energy is proportional to the distance in power law of $\ell^{-3}$ in grand canonical ensemble, whereas it decays as $1/\ell^{7/2}+1/\ell^{11/2}$ in canonical ensemble. Similarly, the Casimir force sharply decreases versus distance between two planar walls as the power law $1/\ell^{9/2}+1/\ell^{13/2}$ in the canonical ensemble instead of $1/\ell^{4}+1/\ell^{7}$ in the grand canonical ensemble \cite{ThuPhysa}. One can easily realize that the power law for the Casimir effect is a half-integer in the canonical ensemble and a integer in the grand canonical ensemble.

Although there are many significant properties, the results in both statistical ensembles have several common points. Firstly, in both the canonical and grand canonical ensembles, the Casimir energy is always negative and it increases monotonously as the distance between two plates increases. This fact leads to the Casimir force is attractive. A further common feature is that the Casimir force is divergent when the distance tends to zero in both statistical ensembles. Last but not least, the Casimir effect does not appear in the ideal Bose gas at zero temperature in the canonical ensemble as well as in the grand canonical ensemble.

It is very interesting to extend this problem to consider the Casimir effect in a binary mixture of Bose gases.

\section*{Acknowledgements}

This research is funded by Vietnam National Foundation for Science and Technology Development (NAFOSTED) under grant number 103.01-2018.02.


\section*{References}


\end{document}